\documentclass[conference]{IEEEtran}
\pdfoutput=1 
\usepackage{epsfig}
\usepackage{tabularx}
\usepackage{graphicx,subfigure} 
\usepackage{color}
\usepackage{xspace}
\usepackage{thumbpdf}
\usepackage{listings}
\usepackage{verbatim}
\usepackage{booktabs}
\usepackage{colortbl}
\usepackage{flushend}

\begin{document}

%%%%%%%%%%%% THIS IS WHERE WE PUT IN THE TITLE AND AUTHORS %%%%%%%%%%%%
\newcommand{\superscript}[1]{\ensuremath{^{\textrm{#1}}}}

\title{The Hayastan Shakarian cut: measuring the impact of network disconnections}

\author{
\IEEEauthorblockN {Ivana Bachmann\IEEEauthorrefmark{1}, Patricio Reyes\IEEEauthorrefmark{2}, Alonso Silva\IEEEauthorrefmark{3}, Javier Bustos-Jim\'enez\IEEEauthorrefmark{1}}\\
\IEEEauthorblockA{\IEEEauthorrefmark{1}NICLabs, Computer Science Department (DCC), Universidad de Chile. Emails: \{ivana,jbustos\}@niclabs.cl}
\IEEEauthorblockA{\IEEEauthorrefmark{2}Technological Institute for Industrial Mathematics (ITMATI), Santiago de Compostela, Spain. Email: patricio.reyes@usc.es}
\IEEEauthorblockA{\IEEEauthorrefmark{3}Bell Labs, Alcatel-Lucent, Centre de Villarceaux, Nozay, France. Email: alonso.silva@alcatel-lucent.com}
}

\maketitle

%\thispagestyle{empty}

%%%%%%%%%%%%%  ABSTRACT GOES HERE %%%%%%%%%%%%%%
\begin{abstract}
In this article we present the \textit{Hayastan Shakarian} (HS), a robustness index for complex networks.
\textit{HS} measures the impact of a network disconnection (edge) while comparing the sizes of the remaining connected components.
Strictly speaking, the \textit{Hayastan Shakarian} index is defined as edge removal that produces the maximal inverse of the size of the largest connected component divided by the sum of the sizes of the remaining ones.

We tested our index in attack strategies where the nodes are disconnected in decreasing order of a specified metric. We considered using the \textit{Hayastan Shakarian} cut (disconnecting the edge with max HS) and other well-known strategies as the higher betweenness centrality disconnection. All strategies were compared regarding the behavior of the robustness ($R$-index) during the attacks. In an attempt to simulate the internet backbone, the attacks were performed in complex networks with power-law degree distributions (scale-free networks).

Preliminary results show that attacks based on disconnecting using the \textit{Hayastan Shakarian cut} are more dangerous (decreasing the robustness) than the same attacks based on other centrality measures. 

We believe that \textit{the Hayastan Shakarian cut}, as well as other measures based on the size of the largest connected component, provides a good addition to other robustness metrics for complex networks.
\end{abstract}

\section{Introduction}
Networks are present in human life in multiple forms from social networks to communication networks (such as the Internet), and they have been widely studied as complex networks of nodes and relationships, describing their structure,  relations,  etc.  Nowadays, studies go deep into network knowledge, and using standard network metrics such as degree distribution and diameter one can determine how robust and/or resilient a network is.

Even though both terms (robustness and resilience) have been used with the same meaning, we will consider \textit{robustness} as the network inner capacity to resist failures, and \textit{resilience} as the ability of a network to resist and recover after such failures. However, both terms have in common the methodology for testing robustness and/or resilience. Usually, they consist on planned attacks against nodes failures or disconnections, from a random set of failures to more elaborated strategies using well known network metrics.  

We consider that an ``adversary'' should plan a greedy strategy aiming to maximize damage with the minimum number of strikes. We  present a new network impact metric called the \textit{Hayastan Shakarian} index (HS), which reflects the size of the larger connected component after removing a network edge. Then, we plan a greedy strategy called the \textit{Hasyastan Shakarian} cut (HSc), which is to remove the edge with higher HS value, aiming to partition the network in same-sized connected components, and recalculating the network \textit{HS}s after each cut. We discuss the performance of attacks based on \textit{the Hayastan Shakarian cut} and the edge betweenness centrality metric \cite{bersano2012metrics}, compared by the robustness index ($R$-index). Our main conclusion is that the first strikes of the strategy   \textit{the Hayastan Shakarian} cut strategy causes more ``\textit{damage}'' than the classical betweenness centrality metric in terms of network disconnection ($R$-index).  

The article is organized as follows, next section presents related work, followed by the definition of  \textit{the Hayastan Shakarian cut}, its attacking strategy, and its simulation results in Section \ref{miuz}.  Conclusions are presented in Section \ref{conclusions}.

\section{Related Work}
\label{related}
Over the last decade, there has been a huge interest in the analysis of complex networks and their connectivity properties~\cite{Albert2000}. During the last years, networks and in particular social networks have gained significant popularity. An in-depth understanding of the graph structure is key to convert data into information. To do so, complex networks tools have emerged~\cite{Albert2002} to classify networks~\cite{Watts1998}, detect communities~\cite{Leskovec2008}, determine important features and measure them~\cite{bersano2012metrics}.

Concerning robustness metrics, betweenness centrality deserves special attention.  Betweenness centrality is a metric that determines the importance of an edge by looking at the paths between all of the pairs of remaining nodes. Betweenness has been studied as a resilience metric for the routing layer~\cite{smith2011network} and also as a robustness metric for complex networks \cite{iyer2013attack} and for internet autonomous systems networks~\cite{mahadevan2006internet} among others.  Betweenness centrality has been widely studied and standardized as a comparison base for robustness metrics, thus in this study it will be used for performance comparison. 

Another interesting  edge-based metrics are: the increase of distances in a network caused by the deletion of vertices and edges \cite{krishnamoorthy1990incremental};  the mean connectivity \cite{tainiter1975statistical}, that is, the probability that a network is disconnected by the random deletion of edges; the average connected distance (the average length of the shortest paths between connected pairs of nodes in the network) and the fragmentation (the decay of a network in terms of the size of its connected components after random edge disconnections) \cite{Albert2000}; the balance-cut resilience, that is, the capacity of a minimum cut such that the two resulting vertex sets contain approximately the same number (similar to the HS-index but aiming to divide the network only in halves \cite{tangmunarunkit2002network}, not in equal sized connected components); the effective diameter (``\textit{the smallest $h$ such that the number of pairs within a $h$-neighborhood is at least $r$ times the total number of reachable pairs}'' \cite{palmer2001connectivity}); and the Dynamic Network Robustness (DYNER) \cite{singer2006dynamic}, where a backup path between nodes after a node disconnection and its length is used as a robustness metric in  communication networks.

The idea of planning a ``network attack'' using centrality measures has captured the attention of researchers and practitioners nowadays. For instance, Sterbenz et al.~\cite{sterbenz2011modelling} used bet\-ween\-ness-centrality (\textit{bcen}) for planning a network attack, calculating the \textit{bcen} value for all nodes, ordering nodes from higher to lower \textit{bcen}, and then attacking (discarding) those nodes in that order. They have shown that disconnecting only two of the top \textit{bcen}-ranked nodes, their packet-delivery ratio is reduced to $60\%$, which corresponds to~$20\%$ more damage than other attacks such as random links or nodes disconnections, tracked by link-centrality and by node degrees.  

A similar approach and results were presented by {\c{C}}etin\-kaya et al.~\cite{ccetinkaya2013modelling}. They show that after disconnecting only $10$ nodes in a network of $100+$ nodes the packet-delivery ratio is reduced to $0\%$. Another approach, presented as an improved network attack \cite{rak2010survivability, sydney2010characterising}, is to recalculate the betweenness-centrality after the removal of each node \cite{holme2002attack,molisz2006end}. They show a similar impact of non-recalculating strategies but discarding sometimes only half of the equivalent nodes. 

In the study of resilience after edge removing, Rosenkratz et al. \cite{rosenkrantz2009resilience} study backup communication paths for network services defining that a network is ``\textit{k-edge-failure-resilient if no matter which subset of k or fewer edges fails, each resulting subnetwork is self-sufficient}'' given that ``\textit{the edge resilience of a network is the largest integer k such that the network is k-edge-failure-resilient}''.

For a better understanding of network attacks and strategies, see  \cite{holme2002attack,molisz2006end,rak2010survivability, sydney2010characterising}.

\section{The Hayastan Shakarian Cut}
\label{miuz}

Given a network $\mathcal{N}$ of size $N$,  we denote by $\mathcal{C}(\mathcal{N} \setminus e)$ the set of connected components in $\mathcal{N}$ after disconnecting edge~$e$. The \textit{Hayastan Shakarian} index (HS-index) for an edge $e$ in $\mathcal{N}$ is defined as follows:
\begin{equation}
{\scriptsize
\textit{HS}_\mathcal{N}(e) = 
  \left\{
    \begin{array}{l}
      \frac{\sum_{\textit{c} \in \mathcal{C} (\mathcal{N}\setminus e)}{\|\textit{c}}\|}{max_{\textit{c} \in \mathcal{C} (\mathcal{N}\setminus e)}\|\textit{c}\|} - 1, \textit{if }  \|\mathcal{C} (\mathcal{N}\setminus e)\| \neq  \|\mathcal{C} (\mathcal{N})\| \\
       \\
      0 \hfill \textit{otherwise}
    \end{array}
  \right.
}
\end{equation}

\noindent where $\|c\|$, with $\textit{c} \in \mathcal{C} (\mathcal{N}\setminus e)$, is the size of the connected component $c$ of the network $\mathcal{N}$ after disconnecting edge $e$. Notice that $\textit{HS}_\mathcal{N}$($e$) reflects the partition of a network in several sub-networks after the disconnection of edge $e$ and how these sub-networks remain interconnected. Strictly speaking, it compares in size the core network (the largest connected component) with the other remaining sub-networks. $\textit{HS}_{\mathcal{N}}$ takes values between $0$ (the whole network remains connected) to $N-1$ (the whole network is disconnected). 

Then, we define that the \textit{Hayastan Shakarian cut} as the one that disconnects the node with the highest \textit{HS-index} value (when there is more than one, we choose it at random). That is:
\[
\textit{HSc} = \mathcal{N}\setminus \hat{e}, \textit{where}~\hat{e} = \textit{max} (\textit{HS}_\mathcal{N}(e)) ~\forall ~e \in \mathcal{N}
\]

\begin{figure}[htp]
       \begin{center}
              \subfigure[Original network]{\includegraphics[width=.55\linewidth]{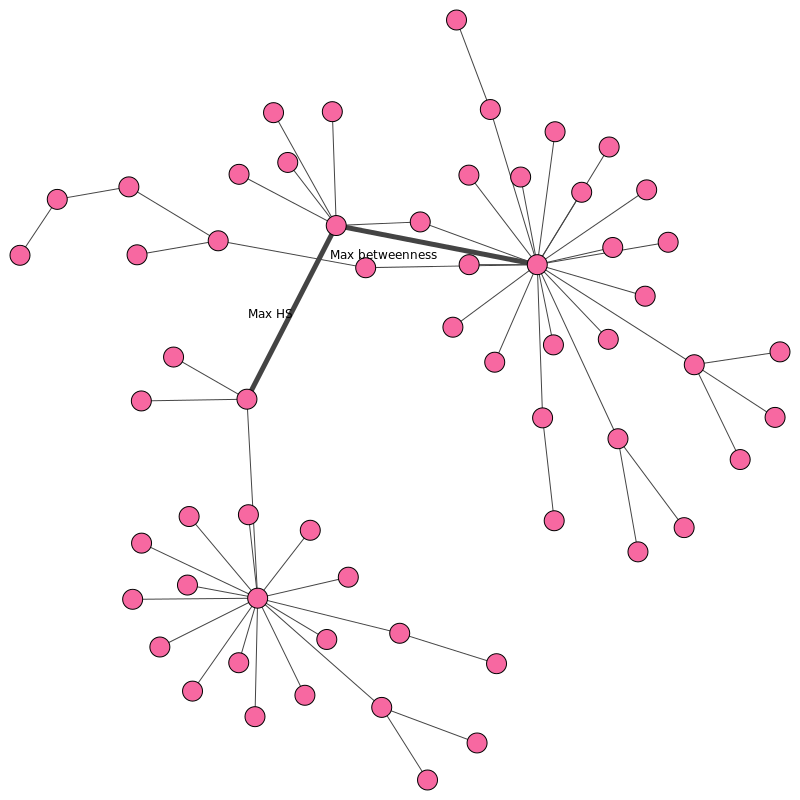}}
              \subfigure[HS removing] {\includegraphics[width=.45\linewidth]{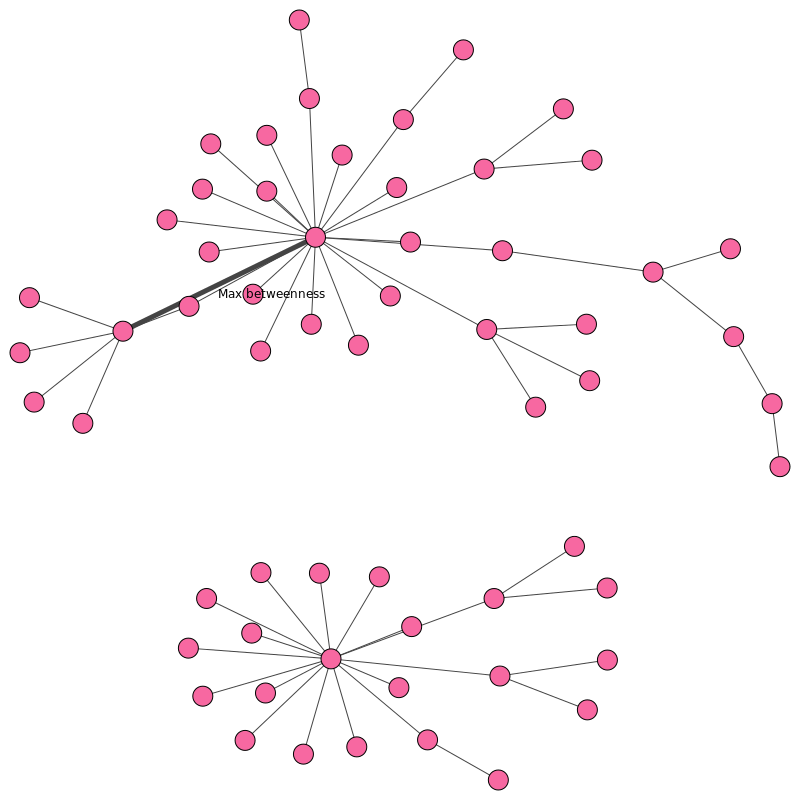}} \hfill
              \subfigure[BC removing]{\includegraphics[width=.45\linewidth]{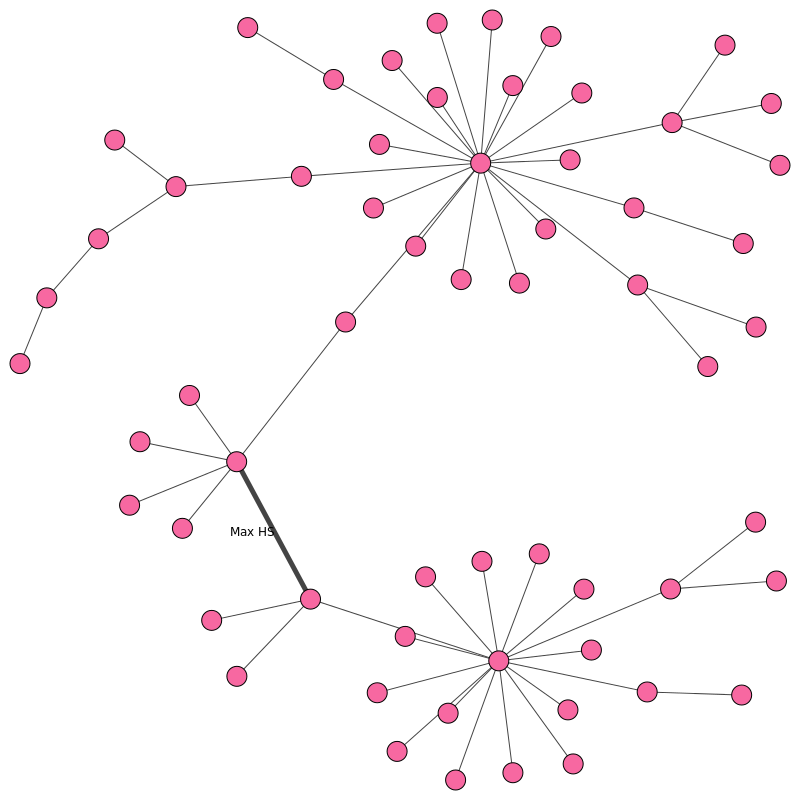}}
       \end{center}
       \caption{Effects of removing edges by (b) its maximal HS, and (c) its maximal betweenness centrality value.}
       \label{fig:figure1}
\end{figure}%

\subsection{Network Attack Plan}

If we plan a network attack by disconnecting nodes with a given strategy (e.g. Figure \ref{fig:figure1}), it is widely accepted to compare it against the use of  betweenness centrality metric, because the latter reflects the importance of an edge in the network \cite{iyer2013attack}. These attack strategies are compared by means of the \textit{Unique Robustness Measure} ($R$-index)~\cite{schneider2011mitigation}, defined as: 
\begin{equation}
R = \frac{1}{N}\sum_{Q=1}^{N} {s(Q)},
\end{equation}
where $N$ is the number of nodes in the network and $s(Q)$ is the fraction of nodes in the largest connected component after disconnecting (remove the edges among connected components) $Q$ nodes using a given strategy.  Therefore, the higher the $R$-index, the better in terms of robustness. Our strategy is:\\

In each step, (re)compute the \textit{HS} index for all the edges, and perform the \textit{Hayastan Shakarian} cut.\\

\subsection{Simulations}

In an attempt to study the behavior of the internet backbone, we test our strategy in simulated power-law degree distributions with exponents $\alpha \in \{1.90,2.00,2.05,2.10,2.20\}$ using Gephi. A power-law distribution is one that can be approximated by the function $f(x) =k*x^{-\alpha}$, where $k$  and $\alpha$ are constants. For each $\alpha$, we simulate $50$ networks of size $1000$ (i.e., with $1000$ nodes).
Then, we tested and compared strategies ranked in terms of the $R$-index. 

Instead of just comparing the robustness, after the removal of all of the nodes, we studied the behavior of the attacks after only a few strikes. To do so, we define a variant of the $R$-index which takes into account only the first $n$ strikes of an attack. Thus, for a simultaneous attack (where the nodes are ranked by a metric only once at the beginning), the $R_n$-index is defined as:
\begin{equation}
R_n = \frac{1}{n}\sum_{Q=1}^{n} {s(Q)}.
\end{equation}
For a sequential attack, the order of node disconnection is recomputed after each disconnection. Similar to the $R$-index, notice that the lower the $R_n$-index, the more effective the attack is, since that gives us a higher reduction of robustness. 

\begin{figure*}[thb!]
       \begin{center}
       \subfigure[$\alpha=1.9$]{\includegraphics[width=.195\linewidth]{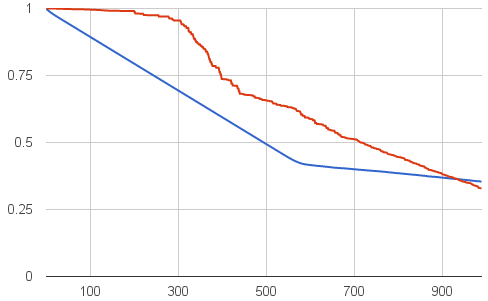}}
\subfigure[$\alpha=2.0$]{\includegraphics[width=.195\linewidth]{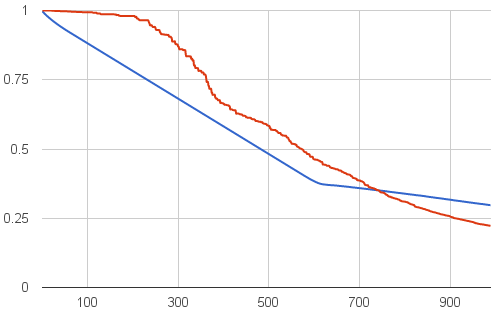}}
\subfigure[$\alpha=2.05$]{\includegraphics[width=.195\linewidth]{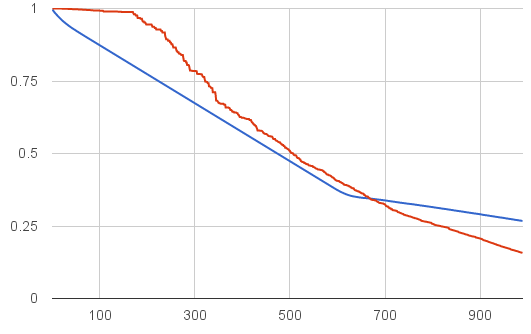}}
\subfigure[$\alpha=2.1$]{\includegraphics[width=.195\linewidth]{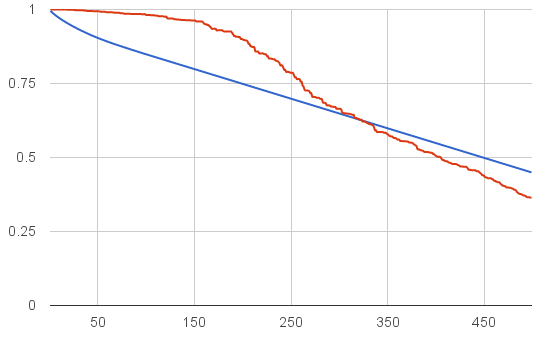}}
\subfigure[$\alpha=2.2$]{\includegraphics[width=.195\linewidth]{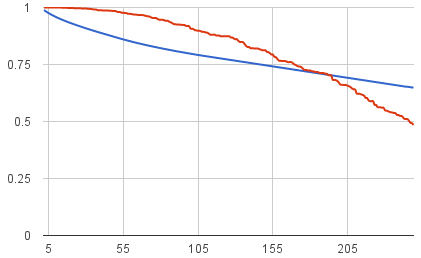}}
       \end{center}
       \caption{Size of the largest fractional connected component (Y-axis) vs. number of disconnected edges (X-axis) using as strategies the Hayastan Shakarian cut (blue) and edge betweenness centrality (red).}
       \label{fig:figureX}
\end{figure*}

Results are shown in Figure~\ref{fig:figureX}. We tested sequential attacks: At each strike, the next node to disconnect was the one with the highest metric (whether it be \textit{HSc} or betweenness centrality) in the current network. The figure shows the behavior of the $R_n$-index in 50 scale-free networks (generated as before) with exponent 1.9 to 2.2. The \textit{Hayastan Shakarian} cut proves to be very effective in attacks with only a few disconnections. Moreover, it shows that the effectiveness persists over the number of strikes in scale-free networks with lower exponent.

It is interesting to note that, no matter the metric used, the damage decreases in the long term with the number of strikes. Revisiting the definition of network robustness, that is, \textit{``a measure of the decrease of network functionality under a sinister attack''}\cite{holme2002attack}, it is important to notice that $R$-index is a metric for a general view of robustness, and it gives no information of how fast the network is disconnected. We suggest that weighted versions of the $R$-index could be a good compromise. This is the case of the $R_n$-index proposed above which takes into account the removal of only a percentage of the best ranked nodes (or until the network reaches its percolation threshold \cite{holme2002attack}) could be a more useful metric for robustness and/or resilience in worst-case scenarios.

We start analyzing what would be the worst ``attack''. In the worst case, a ``\textit{malicious adversary}'' will try to perform the maximum damage with the minimum number of strikes, that is, with the minimum number of node disconnections (made by the attacker).

In terms of decreasing the size of the largest connected component, the \textit{Hayastan Shakarian cut} is the best attack strategy during the first strikes. It achieves the disconnection of more than a quarter of the network only after $200$ strikes for $\alpha=2.10$ and after $150$ strikes for $\alpha=2.05$ (see Figure \ref{fig:figureX}).  It is important to notice that the greater the value of $\alpha$ the lower the number of strikes needed by the \textit{Hayastan Shakarian cut} for reducing the original network to a connected component of $0.75$ of the original size. 

\subsection {Revisiting the Hayastan Shakarian cut}

Concerning the attacking strategies, it is important to notice that the \textit{Hayastan Shakarian cut}  was designed for disconnecting the network from the first strikes, aiming to get connected components with similar sizes. This is the main reason of its better performance compared to the betweenness centrality metric in a worst-attack scenario. For instance, in Figure \ref{fig:figureX} we present  examples where the \textit{HS} cut performs better attacks than an strategy using the betweenness centrality metric for different values of $\alpha$.  Notice that the \textit{Hayastan Shakarian cut} performs better than the edge betweenness centrality strategy for values $\alpha < 2.1$, disconnecting the former more than the half of the network in less strikes than the latter. 

Given that for a high coefficient (of the power law) the network would have a densely connected core with a few nodes, and many nodes of degree $1$. In this case a \textit{Hayastan Shakarian cut} will not be as effective as an edge betweenness  attack since the \textit{HS}-index of the nodes in the core is probably $0$ (because is highly connected), performing a \textit{Hayastan Shakarian cut} over those edges attached to nodes of degree $1$. 

On the other hand, for a low coefficient (of the power law) the proportion of high degree nodes and low degree nodes  is more similar than those of a high coefficient, this means the network is well connected even if it is not densely connected. In this case the \textit{Hayastan Shakarian cut}  will not cause that much damage (because the network is well connected) but it still causes more damage than edge betweenness, because edge betweenness chooses the edges with high traffic, and since the network is well connected it has many paths between any two nodes. Instead the \textit{Hayastan Shakarian cut} will keep choosing the edges that causes more nodes to be disconnected.

Betweenness ``generates'' weak points by removing the edges with high traffic until there are no more alternative paths between 2 nodes. On the other hand the \textit{Hayastan Shakarian cut} finds weak points that are already on the network (before removing a single edge, see Figure: \ref{fig:figure1}). Therefore, the \textit{Hayastan Shakarian cut} will not produce a network necessarily weaker, but a more fragmented and with stronger fragments.

It is important to notice that there is a breaking point where \textit{Hayastan Shakarian cut} is no longer the best attacking strategy. As shown in Figure \ref{fig:figureX}, this breaking point appears after the largest connected component is less than $1/4$ of the original network. While an edge betweenness attack starts attacking the edges in the core with high traffic, this will not produce loss of nodes in the beginning but after a number of edge removals, making the network weak and vulnerable to the point where a single additional edge removal may greatly reduce the fractional size of the largest component, in our examples, of around $0.1$. After the breaking point and for the same number of edges removed by the \textit{Hayastan Shakarian cut} the fractional size of  the largest component will be higher than the one produced by a edge betweenness attack.  

Another interesting property of $\textit{HS}$-index occurs when $\textrm{max} (\textit{HS}_\mathcal{N}(e)) = 0$. In other words, when the \textit{Hayastan Shakarian} index is 0 for any node. In that case, the full network will remain connected no matter which edge is disconnected. Therefore, we suggest a resilience metric as the number (or percentage) of disconnected nodes until $\textrm{max} (\textit{HS}_\mathcal{N}(e)) > 0$, a pre-defined threshold, or the percolation threshold \cite{holme2002attack}.

\section{Conclusions and Future Work}
\label{conclusions}

In this article we have presented the \textit{Hayastan Shakarian cut} (HSc), a robustness index for complex networks defined as the inverse of the size of the remaining largest connected component divided by the sum of the sizes of the remaining connected components.

We tested our index as a measure to quantify the impact of node removal in terms of the network robustness metric $R$-index. We compared \textit{HSc} with other attacks based on well-known centrality measures (betweenness centrality) for node removal selection. The attack strategy used was sequential targeted attack, where every index is recalculated after each removal, and the highest one is selected for the next extraction. 

Preliminary results show that the \textit{Hayastan Shakarian cut} performs better compared to a betweenness centrality-based attack in terms of decreasing the network robustness in the worst-attack scenario (disconnecting more than the half of the network) for values of $\alpha < 2.1$. We suggest that the \textit{Hayastan Shakarian cut}, as well as other measures based on the size of the largest connected component, provides a good addition to other robustness metrics for complex networks.

As future work, we can study improving through new connections an already existing network to make it more robust to attacks, and we can study networks in which the topological space is correlated with the cost of nodes disconnections, nodes which are nearby have a lower cost to be disconnected compared with nodes which are far apart.

\bibliographystyle{abbrv} 
\bibliography{hotnets15}
\label{last-page}

\end{document}